\documentclass[pra,aps,twocolumn,superscriptaddress,floatfix,noshowpacs,nofootinbib,10pt]{revtex4-2}

\usepackage[utf8]{inputenc}     	
\usepackage{amsmath, amssymb, mathtools, mathrsfs}
\usepackage[colorlinks=true]{hyperref}
\hypersetup{colorlinks,linkcolor={red},citecolor={blue},urlcolor={blue}}  
\usepackage{color}
\usepackage{graphicx}
\usepackage[normalem]{ulem}
\usepackage{mathtools}
\usepackage{stmaryrd}
\usepackage{float}
\usepackage{textalpha}

\newcommand{\bs}[1]{{\boldsymbol{#1}}}
\newcommand{\bk}{\bs{k}}
\newcommand{\br}{\bs{r}}

\newcommand{\bq}{\bs{q}}
\newcommand{\bp}{\bs{p}}

\begin{document}
\title{Critical non-thermal fixed point and the dynamical condensation phase transition}
\author{Anne-Sol\`ene Bornens}
\email{annesolene.bornens@lkb.upmc.fr}
\affiliation{Laboratoire Kastler Brossel, Sorbonne Universit\'{e}, CNRS, ENS-PSL Research University, 
Coll\`{e}ge de France; 4 Place Jussieu, 75005 Paris, France}

\author{Elisabeth Gliott}
\affiliation{Universit\'e Paris-Saclay, CNRS, Ecole Normale Sup\'erieure Paris-Saclay, CentraleSup\'elec, LuMIn, Orsay, France}

\author{Nicolas Cherroret}
\email{nicolas.cherroret@lkb.upmc.fr}
\affiliation{Laboratoire Kastler Brossel, Sorbonne Universit\'{e}, CNRS, ENS-PSL Research University, 
Coll\`{e}ge de France; 4 Place Jussieu, 75005 Paris, France}

\begin{abstract}
Using a non-perturbative quantum kinetic framework, we develop a unified description of the far-from-equilibrium dynamics of three-dimensional Bose gases following cooling quenches across the Bose-Einstein condensation transition. By tracking the spatio-temporal evolution of the momentum distribution, we show that the equilibrium condensation threshold simultaneously acts as a dynamical critical point, separating distinct far-from-equilibrium universality classes governed by different non-equilibrium attractors.
While quenches above the transition exhibit a single-timescale relaxation toward a thermal fixed point, quenches below the transition display a crossover from a transient weak-turbulence regime to a coarsening fixed point governed by the diffusive recombination of vortex lines. Quenches directly to the condensation threshold, finally, are controlled by a previously unidentified critical fixed point characterized by the superdiffusive spreading of critical fluctuations and a distinct set of dynamical exponents. Together, these dynamical scaling laws establish a far-from-equilibrium counterpart of the condensation phase transition, in which the equilibrium critical point also organizes the long-time non-equilibrium dynamics.
\end{abstract}
\maketitle

The emergence of universal behavior far from equilibrium is a central theme of modern many-body physics \cite{Hohenberg1977, Polkovnikov2011}. While equilibrium phase transitions are understood through the concepts of scaling, universality and criticality, it remains an open challenge to determine whether analogous organizing principles govern the dynamics of strongly driven or quenched systems. Yet a wide range of non-equilibrium phenomena, including turbulence \cite{Nazarenko2011}, coarsening dynamics \cite{Cugliandolo2015} and the formation of ordered phases \cite{Bray2002} following a quench, displays robust large-scale behavior that appears largely insensitive to microscopic details. Understanding whether such phenomena can be classified into a small number of universal dynamical regimes, and how these regimes are connected, is a central goal of non-equilibrium statistical mechanics.

A key concept that has emerged in this context is dynamic scaling, whereby correlations evolve self-similarly in space and time according to universal scaling laws. Such behavior is often associated with non-thermal fixed points (NTFPs), which act as attractors of the far-from-equilibrium dynamics \cite{Berges2008}. Dynamic scaling has been extensively studied in the context of isolated quantum gases following a quench \cite{Nowak2012, Comaron2019, Chantesana2019, Mikheev2019, Zhu2023, Gliott2024}, suggesting that universality does extend far beyond thermal equilibrium. 
In parallel, the study of far-from-equilibrium systems has stimulated renewed interest in the notion of \emph{dynamical phase transition} (DPT), which separates  long-time dynamical regimes associated with different non-equilibrium fixed points \cite{Marino2022}. Such transitions have notably been predicted in formal quenched  $N$-component bosonic \cite{Chandran2013, Sciolla2013, Smacchia2015, Diehl2017, Cherroret2024, delPozo2026} and fermionic \cite{Jian2019, Jian2021} field-theoretic models, where varying the initial conditions can drive the system across a non-equilibrium critical point separating qualitatively different asymptotic dynamics of correlations. 
A natural open question is whether equilibrium critical points in realistic interacting systems can play this dual role, organizing not only thermodynamic phases but also the universal long-time dynamical behavior following a quench.

Ultracold quantum gases provide an ideal platform to address these questions \cite{Langen2015}. In these systems, external driving or quantum quenches can generate phenomena such as turbulent transport \cite{Navon2019, Galka2022, Bagnato2022},  topological-defect formation \cite{Beugnon2017, Keim2017}, or prethermalization and coherence growth \cite{Kollar2011, Gring2012, Natu2013, Larre2018, Pietraszewicz2019, Mallayya2019, Bardon-brun2020}, often long before thermalization eventually sets in \cite{Buchhold2016, Regemortel2018, Bertini2015, Mori2018, Duval2023, Duval2025}. In particular, over the past decade, a series of experiments has investigated quenches across phase transitions in  isolated gases, including Ising-like transitions in spinor Bose gases \cite{Huh2024, Manovitz2025}, as well as thermodynamic transitions such as the Bose-Einstein condensation (BEC) \cite{Erne2018, Glidden2021} and the Kosterlitz-Thouless transitions \cite{Sunami2022, Abuzarli2022, Gazo2023}, alongside extensive theoretical work \cite{Nowak2012, Comaron2019, Chantesana2019, Mikheev2019, Scoquart2022, Gliott2025}. Despite the identification of several dynamical regimes in these settings, a unified picture of far-from-equilibrium quench dynamics across phase transitions remains elusive. In particular, it is not clear whether quenches terminating above, below, or precisely at a critical threshold are governed by distinct attractors. If so, the equilibrium transition would simultaneously define a dynamical critical point, controlled by the post-quench energy rather than the temperature and separating different non-equilibrium scaling regimes.

\begin{figure}[h!]
\includegraphics[scale=0.43]{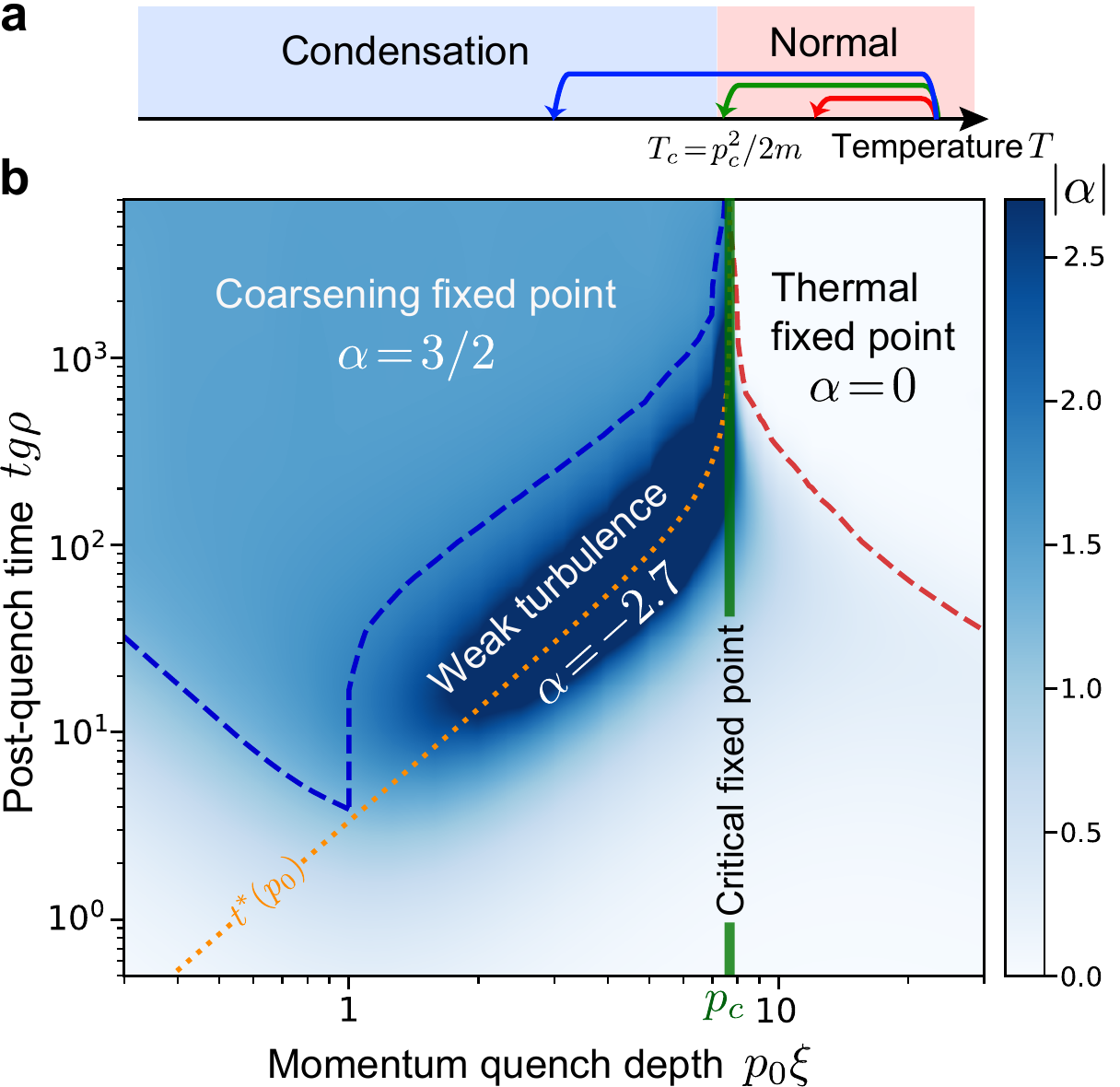}
\caption{\label{Fig:phasediag}
\textbf{a}. Equilibrium phase diagram of the BEC transition, showing the normal and condensed phases separated by the critical temperature $T_c=p_c^2/2m$. The arrows indicate the cooling-quench protocols leading to the DPT investigated in this work.
\textbf{b}. Corresponding dynamical phase diagram for cooling quenches from the normal phase. The color map shows the absolute instantaneous exponent $|\alpha(t)|=|d\ln n_0/d\ln t|$, i.e., the local slope of the curves in Fig.~\ref{Fig:n0}, which characterizes the algebraic growth of the zero-momentum occupation $n_0(t)$n as a function of post-quench time and momentum $p_0$.
The DPT organizes the long-time dynamics into three universal regimes: for $p_0>p_c$, the system relaxes to a thermal fixed point [$\smash{n_0(t\to\infty)\sim t^0}$]; for $p_0<p_c$, it approaches a coarsening NTFP [$\smash{n_0(t\to\infty)\sim t^{3/2}}$]; for $p_0=p_c$, the dynamics is governed by a distinct critical NTFP discussed in the text. For weak quenches below threshold, $1/\xi\leq p_0<p_c$,  a transient weak-turbulence regime emerges, characterized by $n_0(t)\sim |t-t^*|^{-2.7}$ around  the crossover time  $t^*$ (dotted curve) where the Boltzmann description breaks down.  The blue (red) dashed curve marks the time at which the value $\alpha=3/2$ ($\alpha=0$) is reached to within $90\%$. 
}
\end{figure}

Here, we show that the BEC transition organizes not only equilibrium phases but also the universal long-time dynamics following a cooling quench. This is illustrated by the dynamical phase diagram in Fig.~\ref{Fig:phasediag}, which summarizes the dynamics of three-dimensional (3D) Bose gases quenched across the condensation threshold. Using a non-perturbative quantum kinetic theory, we show that while quenches above the threshold relax directly to thermal  equilibrium,  quenches to and below the threshold give rise to distinct NTFPs. Specifically, quenches below the  threshold exhibit two successive non-equilibrium stages: the system first enters a transient weak-turbulence regime, characterized by weak nonlinear interactions of the bosonic field, before crossing over to a long-time \emph{coarsening fixed point} governed by the diffusive recombination of vortex lines.
 To our knowledge, this crossover between two qualitatively distinct non-equilibrium regimes has not been identified previously. Most strikingly, quenches directly to the condensation threshold are controlled by a separate \emph{critical fixed point}, characterized by the superdiffusive spreading of critical fluctuations and a previously unknown set of dynamical scaling exponents.
Altogether, these results establish the concept of a dynamical condensation transition, whereby tuning the post-quench energy across the BEC threshold drives the system toward distinct non-equilibrium fixed points governing different asymptotic scaling regimes. 

\section*{Model}

\begin{figure}
\includegraphics[scale=0.75]{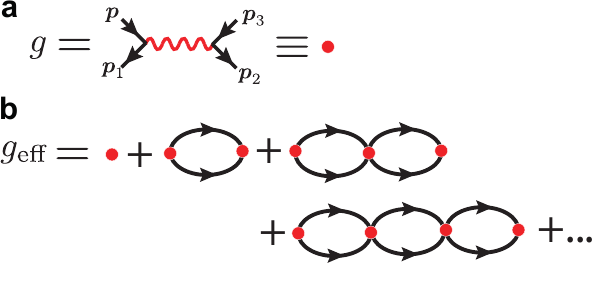}
\caption{\label{Fig:diagrams}
\textbf{a}. Zeroth-order diagrammatic representation of the contact interaction $g$, leading to the Boltzmann equation. 
\textbf{b}. Self-consistent renormalization of the contact interaction, accounting for repeated scattering events between two particles and leading to the effective interaction \eqref{eq:geff}.}
\end{figure}
We consider a uniform Bose gas of $N$ particles of mass $m$ confined in a 3D box of size $L$. We work in the thermodynamic limit $N,L\to\infty$ and in the dilute regime $\rho a^3\ll~1$, where $\rho\equiv N/L^3$ is the (constant) mean particle density and $a$ the $s-$wave scattering length characterizing contact interactions. The diluteness condition can equivalently be expressed as $\rho\xi^3\gg1$, where $\xi\equiv 1/\sqrt{2mg\rho}$ is the healing length, with $g\equiv 4\pi a/m$ (hereafter we set $\hbar=1$). In all subsequent numerical calculations, we fix the dimensionless parameter $4\pi^2\rho\xi^3=10^3$.
The gas is initially assumed to be in thermal equilibrium in its  normal phase, i.e., at a temperature $T>T_c$ above the critical temperature $T_c=2\pi/m[\rho/\zeta(3/2)]^{2/3}$ of the BEC transition \cite{Stringari_pitaevskii2003}. At  time $t=0$, we suppose that the temperature is suddenly reduced, as illustrated in Fig. \ref{Fig:phasediag}a. In practice, this can be implemented through a rapid evaporative cooling protocol, as experimentally realized, e.g., in \cite{Glidden2021}. Following the quench, the Bose gas is driven out of equilibrium and undergoes a  redistribution of the occupation numbers $n_\bp(t)$ of the momentum modes $\bp$.
We describe this dynamics through the quantum kinetic equation \cite{Chantesana2019, Mikheev2019}
\begin{align}
\label{Boltzmann_eq}
&\partial_t n_\bp={4\pi}\int d^3\bp_1 d^3\bp_2 d^3\bp_3\, g^2_{\text{eff}}({\epsilon_\bp\!-\! \epsilon_{\bp_2}}, \bp\!-\!\bp_2 ,t) \nonumber\\
&\ \ \ \ \times\delta(\bp_1\!+\!\bp_2\!-\!\bp\!-\!\bp_3)
\delta(\epsilon_{\bp_1}\!+\!\epsilon_{\bp_2}\!-\!\epsilon_{\bp}\!-\!\epsilon_{\bp_3})\nonumber\\
&\ \ \ \ \times \left[n_{\bp_1}n_{\bp_2}(1\!+\!n_{\bp}\!+\!n_{\bp_3})\!-\!n_{\bp}n_{\bp_3}(1\!+\!n_{\bp_1}\!+\!n_{\bp_2})\right],
\end{align}
where $\epsilon_{\bp}\equiv \bp^2/2m$.
The collision integral on the right-hand side describes the rate of change of the population of mode $\bp$ due to  repulsive  boson interactions.
It accounts for two-body scattering processes that either deplete mode $\bp$ via $(\bp,\bp_3)\to(\bp_1,\bp_2)$ or populate it through the reverse process $(\bp_1,\bp_2)\to(\bp,\bp_3)$. These scattering events are constrained by momentum and energy conservation, encoded in the two Dirac delta functions, and weighted by the squared effective coupling strength $g^2_\text{eff}(\epsilon,\bq,t)$, which generally depends on the momentum transfer $\bq$ and energy transfer $\epsilon$ of the scattering process. Importantly, Eq.~(\ref{Boltzmann_eq}) conserves both the total 
particle density, $\rho\equiv \int d^3\bp/(2\pi)^3 n_\bp(t)$, and the total energy density, $e\equiv \int d^3\bp/(2\pi)^3 \epsilon_\bp n_\bp(t)$. 
At leading-order in perturbation theory, $g_\text{eff}=g$ reduces to the bare contact interaction strength in three dimensions \cite{Stringari_pitaevskii2003}, represented diagrammatically in Fig.~\ref{Fig:diagrams}a. 
Within this approximation, Eq. \eqref{Boltzmann_eq} reduces to the standard Boltzmann equation \cite{Griffin2009}, also known as the weak-wave turbulence equation in the regime of large occupation numbers \cite{Nazarenko2011}. Although this approximation  captures several aspects of the condensation dynamics, it is also known to develop a finite-time singularity at a characteristic time $t^*$, preventing  a reliable description of the long-time evolution  \cite{Semikoz1995, Semikoz1997, Berloff2002, Josserand2006, Semisalov2021, Zhu2023}.
To overcome this limitation, we consider a non-perturbative effective interaction $g_\text{eff}$, first introduced in Ref.~\cite{Chantesana2019} and represented diagrammatically in Fig.~\ref{Fig:diagrams}b, which physically accounts for the possibility that two particles  interact recurrently during their evolution. As shown below, this approach regularizes the singular behavior of the Boltzmann equation and enables the description of  NTFPs emerging after quenches both below and exactly at the critical point. The non-perturbative effective coupling reads \cite{Chantesana2019}
\begin{align}
\begin{aligned}
    g_{\text{eff}}(\epsilon, \bp , t) = \frac{g}{\left| 1+ g \Pi(\epsilon, \bp,t)\right|},
    \end{aligned}
\label{eq:geff}
\end{align}
and involves the renormalization factor $\Pi(\epsilon, \bp,t)= \int d^3\bk/(2\pi)^3 n_{\bk}(t)[G(\epsilon,\epsilon_\bk-\epsilon_{\bp-\bk})-G(\epsilon,\epsilon_{\bp-\bk}-\epsilon_\bk)]$, where $G(\epsilon,\epsilon_{\bq})\equiv(\epsilon-\epsilon_{\bq}+i0^+)^{-1}$. 
The physical significance of this renormalization can be qualitatively understood as follows. After a quench below $T_c$, the gas progressively develops long-range phase coherence, leading to a strong accumulation of particles at low momenta through an inverse particle cascade. As the occupations of these modes grow, recurrent scattering processes increasingly renormalize the interaction strength, causing the Boltzmann approximation $g_{\rm eff}\simeq g$ to break down.

To investigate the quench dynamics around $T_c$, we solve Eq.~\eqref{Boltzmann_eq} supplemented with an initial condition $n_\bp(t=0)$, which physically represents the momentum distribution  of the  gas immediately after the quench. In experiments, such quenches are typically implemented by suppressing the high-momentum tail of an equilibrium distribution beyond a characteristic momentum scale $p_0$, which quantifies the quench strength. Throughout this work, we choose $n_\bp(0)\propto \exp(-p^4/p_0^4)$, but the specific  functional form of this distribution  does not affect  the universal dynamics discussed below. The latter are entirely controlled by the post-quench energy density, $e= \int d^3\bp/(2\pi)^3 \epsilon_\bp n_\bp(0)  \propto \rho\,{p_0^2}/2m$, which remains conserved throughout the subsequent evolution. 

\begin{figure}
\includegraphics[scale=0.4]{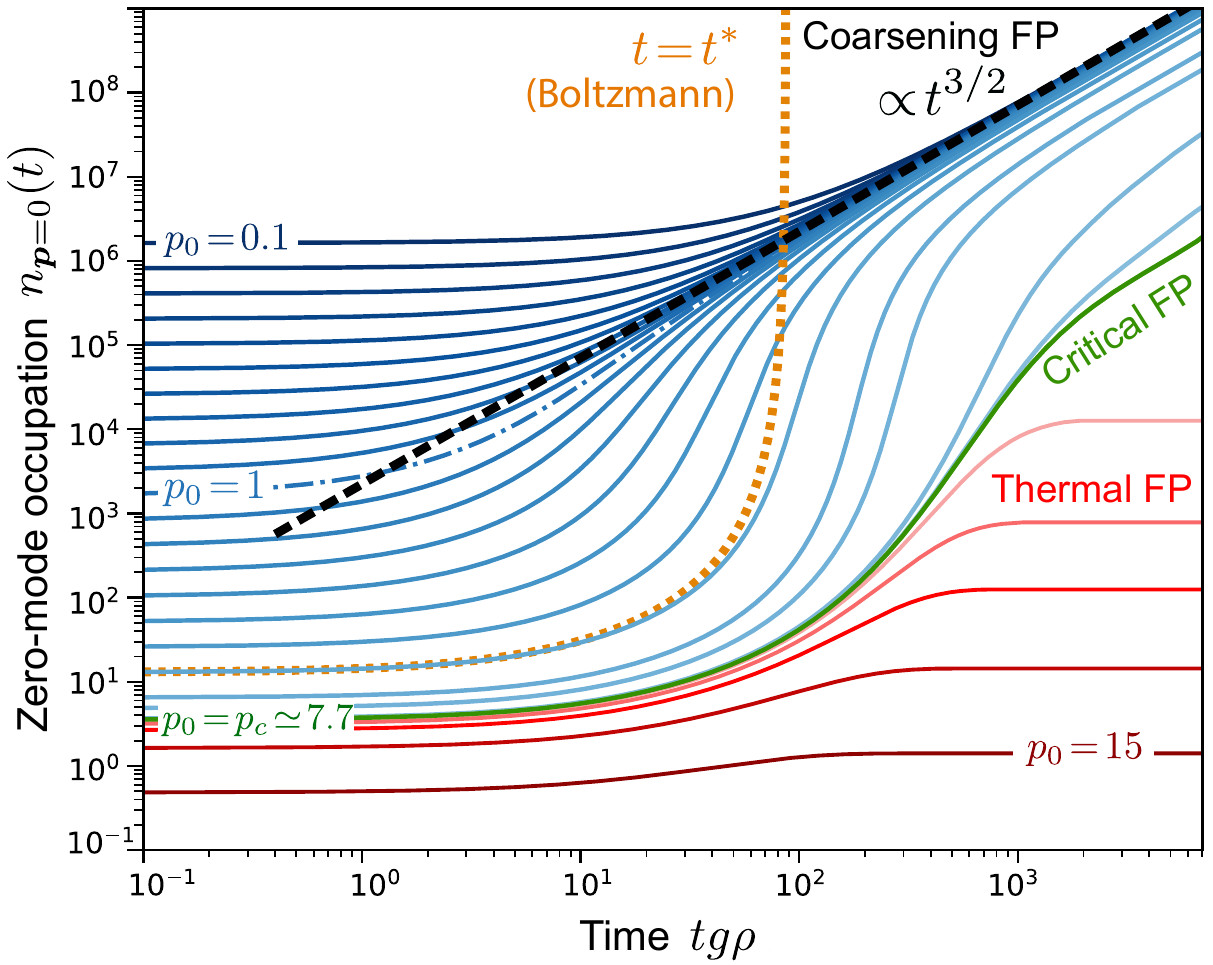}
\caption{\label{Fig:n0}
Zero-mode population $n_0(t)$ of the Bose gas as a function of time for increasing values of the post-quench momentum $p_0$, obtained by numerically solving  Eq.~\eqref{Boltzmann_eq}.
For quenches above the condensation threshold ($p_0>p_c$, red curves), $n_0(t)$ saturates to the thermal value $\sim T/|\mu|$, with $T\to T_c$ and $\sqrt{|\mu|}\propto(p_0-p_c)$ as $p_0\to p_c^+$, see Eq.~\eqref{eq:critical_xi}.
For quenches below or to the threshold ($p_0\leq p_c$), the dynamics asymptotically approaches the universal scaling law $n_0(t)\propto t^{3/2}$. Weak quenches below the threshold ($1/\xi\leq p_0<p_c$) exhibit an intermediate weak-turbulence regime with $n_0(t)\propto |t-t^*|^{\alpha}$ and $\alpha\simeq -2.7$, before crossing over to a long-time coarsening fixed point (FP) with $\alpha=3/2$. The transient regime occurs around the characteristic time $t^*(p_0)$, where the solution of the Boltzmann equation develops a divergence. This solution is shown as a dotted curve for $p_0=5/\xi$.
For deep quenches ($p_0<1/\xi$) driving the gas close to its ground state, the weak-turbulence regime disappears and the dynamics reaches the long-time fixed point directly.
In the figure, $p_0$ is expressed in units of $1/\xi$.
}
\end{figure}
We first focus on  the time evolution of the zero-mode population $n_{\bp=0}(t)$  by numerically solving  Eq.~\eqref{Boltzmann_eq} (see Appendix \ref{app:num} for details about the numerical resolution). The results, shown in Fig. \ref{Fig:n0} for different values of the quench parameter $p_0$, reveal qualitatively distinct dynamical behaviors depending  on whether  $p_0$ is larger, smaller or equal to a critical value $p_c=\sqrt{2m T_c}$, which coincides with the critical momentum of the BEC transition. For $p_0>p_c$, corresponding to a quench that leaves the  gas in the normal phase, $n_{0}(t)$ approaches a finite stationary value at long times, indicating a fast thermalization process. By contrast, for  $p_0< p_c$, $n_{0}(t)$ exhibits an algebraic  growth in time, signaling an evolution toward condensation. As we will see below, the case of a quench  exactly at the critical point, $p_0= p_c$, exhibits a distinctive scaling dynamics. These  three different dynamical regimes define the far-from-equilibrium counterpart of the condensation phase transition, which is summarized by the dynamical phase diagram in Fig. \ref{Fig:phasediag}b and that we now discuss in detail. 

\section*{Thermal fixed point}

\begin{figure*}
\includegraphics[scale=1.25]{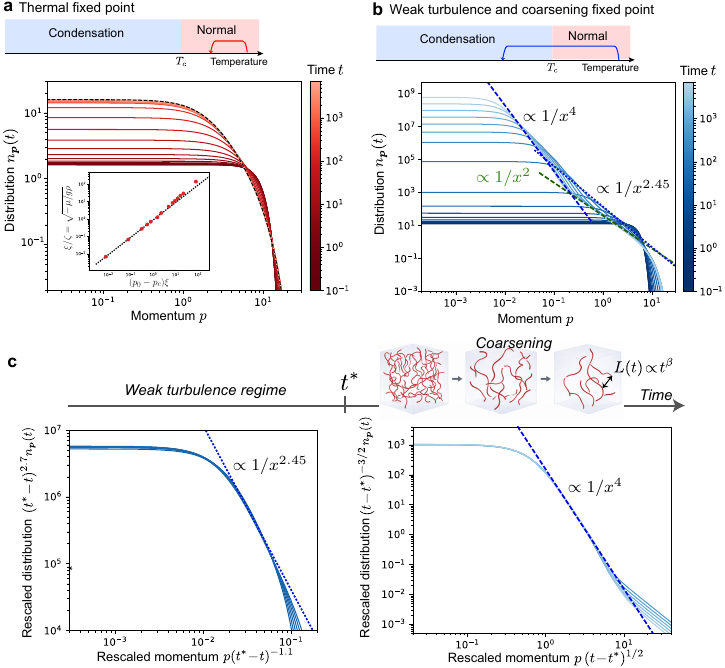}
\caption{
\label{Fig:distributions}
\textbf{a}. Momentum distribution  $n_\bp(t)$ at different times for $p_0=10/\xi>p_c$, corresponding to a quench within the normal phase of the Bose gas.  The lowest (darkest) curve shows the initial distribution, $n_\bp(t)\propto \exp(-p^4/p_0^4)$, while the upper dashed curve is the asymptotic thermal distribution $n^\text{th}_p=[\exp((\epsilon_p-\mu)/T)-1]^{-1}$, with $T=72.6g\rho$ and $\mu=-4.43g\rho$. 
\textbf{b}. Same as in panel a, but for $p_0=5/\xi$, corresponding to a quench across the BEC transition. At intermediate momenta, $n_\bp(t)\sim p^{-\gamma}$ exhibits an algebraic decay, with $\gamma\sim 2.45$ at short times (weak-turbulence regime) and $\gamma\sim 4$ at long times (coarsening fixed point). 
At long times, the distribution tail also develops the thermal scaling $n_\bp(t)\sim p^{-2}$, signaling the rethermalization of non-condensed particles.
\textbf{c}. Momentum distributions at different times, rescaled according to Eq. \eqref{eq:WT_scaling} using $(\alpha,\beta)=(-2.7,-1.1)$ in the weak-turbulence regime ($t\simeq t^*$, left panel), and $(\alpha,\beta)=(3/2,1/2)$ in the coarsening regime ($t\gg t^*$, right panel) \cite{footnote1}. The crossover time $t^*\simeq 86.9/g\rho$ is defined as the time at which the kinetic equation \eqref{Boltzmann_eq} diverges in the perturbative limit $g_\text{eff}=g$. The rescaled distributions are computed from times within the intervals $[30/g\rho,200/g\rho]$ (left panel) and $[500/g\rho,7000/g\rho]$ (right panel). In all graphs, momenta are in unit of $1/\xi$ and times in unit of $1/g\rho$.
}
\end{figure*}

Let us first consider quenches with  $p_0>p_c$ (red solid curves in Fig. \ref{Fig:n0}),  for which $n_0(t)$ saturates at long times. 
This scenario corresponds to a rethermalization of the Bose gas  within the normal phase, so that $n_0(t)$ is expected to converge toward a regular thermal distribution. 
This  is confirmed in Fig. \ref{Fig:distributions}a, which shows the time evolution of the  momentum distribution $n_\bp(t)$ for $p_0=10/\xi$. As $t\to\infty$, $n_\bp(t)$ becomes time independent and  coincides with  the thermal  solution 
$n^\text{th}_p=[\exp((\epsilon_p -\mu)/T)-1]^{-1}$ of Eq.~\eqref{Boltzmann_eq}. We refer to this asymptotic state as the \emph{thermal fixed point} of the DPT. It is fully characterized by two Lagrange multipliers,  the temperature $T$ and the chemical potential $\mu$, whose values are uniquely determined by particle and energy conservation.
As the quench strength approaches the critical value from above, $p_0\to p_c^+$, we find that the long-time occupation of the zero-momentum mode diverges, $n_0(t)\simeq T/|\mu|$, with $T\simeq T_c$ and 
\begin{align}
\label{eq:critical_xi}
\sqrt{-2m\mu}\equiv \zeta^{-1}\underset{p_0\to p_c^+}{\simeq}\frac{3\zeta(3/2)}{5\sqrt{\pi}}(p_0-p_c)^\nu
\end{align}
with $\nu=1$. Here $\zeta$ is nothing but the equilibrium correlation length of the Bose gas above the BEC transition \cite{Smith2017}. The long-time dynamics above the critical point therefore reproduces the conventional equilibrium critical behavior of the 3D Bose gas in its normal phase.
The inset of Fig.~\ref{Fig:distributions}a compares the numerical values of $\zeta$, extracted from the long-time limit of $n_0(t)$, with the analytical prediction \eqref{eq:critical_xi} as a function of the quench parameter $p_0$ close to $p_c$ (red points and dotted line, respectively).

\section*{Weak turbulence and coarsening fixed point}

We now turn to quenches across the BEC transition, $p_0<p_c$, for which an inverse cascade develops.
The corresponding  evolution of the zero-mode population is shown by the blue curves in Fig.~\ref{Fig:n0}. A rich dynamical scenario emerges for moderate quenches, $1/\xi\leq p_0<p_c$, corresponding to a post-quench energy per particle $e/\rho$  larger than the ground-state energy $\sim g\rho$. 
In this regime, the dynamics proceeds through two successive stages. First,  $n_0(t)$  grows rapidly until an inflection point around a characteristic time $t\sim t^*$. It then crosses over to a purely algebraic growth at longer times. These two stages are also reflected in the momentum distribution $n_\bp(t)$ shown in Fig. \ref{Fig:distributions}b. At early times, the distribution displays an intermediate-momentum power-law behavior $n_\bp(t)\sim 1/p^{2.45}$, whereas  at long times it evolves toward the steeper scaling $n_\bp(t)\sim 1/p^{4}$. 

To understand the origin of these two dynamical regimes, we insert the generic scaling ansatz  $n_\bp(t)=\tau(t) f(p\,\tau(t)^y)$ into Eq. \eqref{Boltzmann_eq}, assuming that the effective coupling strength scales as $g_\text{eff}(\epsilon_\bp\!-\!\epsilon_{\bp_2},\bp\!-\!\bp_2)\sim |\bp-\bp_2|^z$ at low momentum. Separating variables then yields the differential relation $\tau'(t)\tau(t)^{2(2+z)y-3}=\text{const}$, from which it follows that $\tau(t)\propto |t^*-t|^\alpha$, with $\alpha={1/(2y(2+z)-2)}$. Here, $t^*$ arises as an  integration constant defined by $\tau(t^*)=0$. Introducing $\beta=y\alpha$, we finally obtain:
\begin{equation}
\label{eq:WT_scaling}
n_\bp(t)=|t^*\!-\!t|^\alpha f\big(p|t^*\!-\!t|^\beta\big),
\end{equation}
together with the exponent relation
\begin{equation}
\label{eq:alphabeta_relation}
(2+z)\beta-\alpha=\frac{1}{2}.
\end{equation}
As we show below, Eq.~\eqref{eq:WT_scaling} encompasses the two regimes identified numerically above. Both exhibit an intermediate momentum window in which the scaling function behaves as $f(x)\sim x^{-\gamma}$, but they are characterized by two different sets of dynamical exponents $(\alpha,\beta,\gamma)$.

The first scaling regime emerges around the characteristic time $t\sim t^*$, where the momentum distributions indeed obey Eq.~\eqref{eq:WT_scaling}. As shown in the left panel of Fig.~\ref{Fig:distributions}c, distributions at successive times collapse onto a single scaling function for $\alpha\simeq-2.7$ and $\beta\simeq-1.1$, in agreement with the exponent relation \eqref{eq:alphabeta_relation} for $z=0$.
We refer to this transient phase as the \emph{weak-turbulence regime}. It corresponds to the universal far-from-equilibrium state reached while the dynamics remains governed by weak nonlinear waves, that is, as long as $g_\mathrm{eff}\simeq g$.
This scaling regime was previously identified in studies based on the Boltzmann equation \cite{Semikoz1995, Semikoz1997, Berloff2002, Josserand2006, Semisalov2021, Zhu2023}, 
with the notable difference that imposing $g_\mathrm{eff}=g$ exactly leads to a finite-time divergence at $t=t^*$,  illustrated by the yellow dotted curve in Fig.~\ref{Fig:n0}. 
While this singularity has long prevented a quantitative description of cooling quenches in Bose gases \cite{Semikoz1997}, the non-perturbative framework regularizes the divergence and captures the dynamics both near and beyond $t^*$.
As further shown in the left panel of Fig. \ref{Fig:distributions}c, the corresponding scaling function  decays algebraically  at intermediate momenta, 
$f(x)\sim 1/x^{\gamma}$, with $\gamma=\alpha/\beta\simeq 2.45$, a  property that  implies a \emph{self-similar} evolution of $n_\bp(t)$.

At times $t\gg t^*$ where the Boltzmann approximation breaks down, Fig.~\ref{Fig:n0} reveals the emergence of a new non-equilibrium regime characterized by $n_0(t)\propto t^\alpha$, with $\alpha\simeq 1.5$. 
The  right panel of Fig. \ref{Fig:distributions}c shows the corresponding long-time momentum distributions, rescaled according to Eq.~\eqref{eq:WT_scaling}, but now using  $\alpha\simeq 1.5$ and $\beta\simeq 0.5$ \cite{footnote1}. 
The resulting collapse demonstrates that the system has reached an asymptotic NTFP, which we refer to as the \emph{coarsening fixed point}, a terminology justified below. 
At this fixed point, $\alpha$ and $\beta$ still satisfy the relation (\ref{eq:alphabeta_relation}), but now with $z=2$.
 The corresponding low-momentum behavior  of the effective coupling, $g_\text{eff}(p)\sim p^2$, was first identified in Ref. \cite{Chantesana2019}, and is analyzed in more detail in Appendix \ref{app:geff}. 
The exponents at the coarsening fixed point additionally obey the hyperscaling relation $\alpha=d\beta$ (with $d=3$ the spatial dimension), which follows from particle-number conservation, $\rho=\int d^3\bp/(2\pi)^3\,  n_\bp(t)$, assuming that the scaling form \eqref{eq:WT_scaling} extends over a sufficiently broad momentum range \cite{Chantesana2019}. 
\begin{figure*}
\includegraphics[scale=0.72]{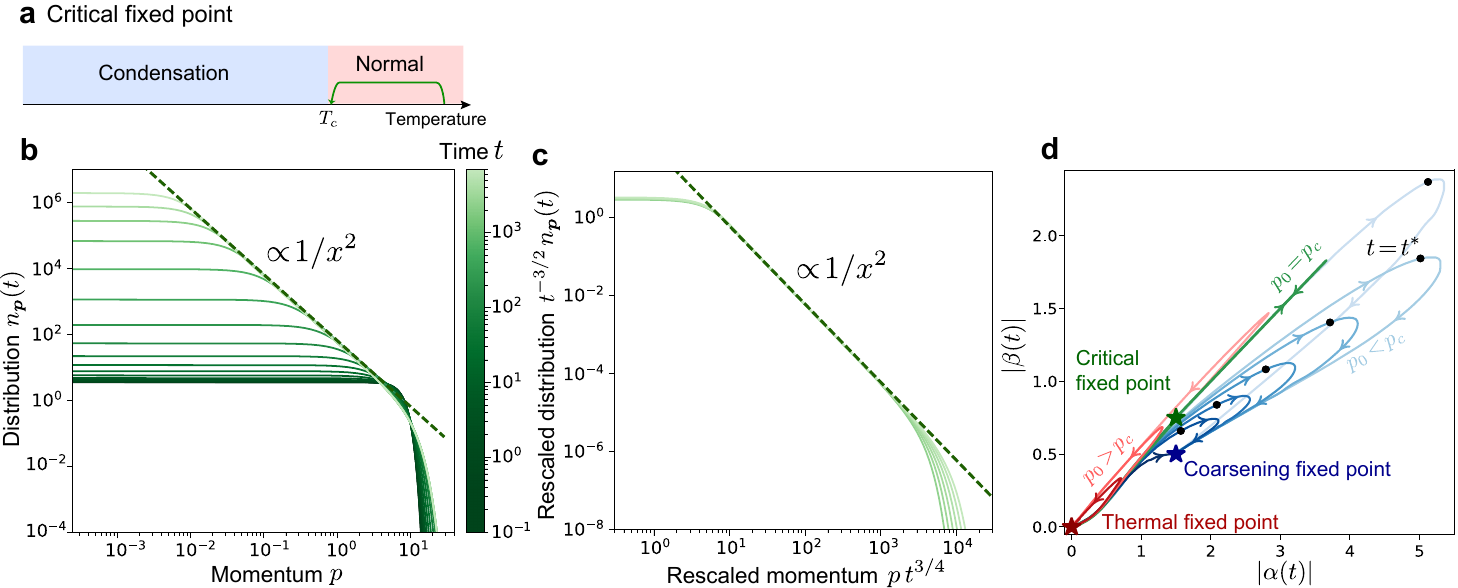}
\caption{
\label{Fig:distributionsCFP}
\textbf{a}. Illustration of a quench to the condensation threshold, $p_0=p_c$. 
\textbf{b} Momentum distribution $n_\bp(t)$ at increasing times for $p_0=p_c=7.7/\xi$. At intermediate momenta, $n_\bp(t)\sim p^{-\gamma}$ with $\gamma=2$. 
\textbf{c}. Rescaled distributions  according to Eq.~\eqref{eq:WT_scaling}, using $(\alpha,\beta)=(3/2,3/4)$ and $t^*=0$. The collapse is shown for times $t>2500/g\rho$. In \textbf{b} and \textbf{c}, momenta are expressed in units of $1/\xi$ and times in units of $1/g\rho$. 
\textbf{d} ``Flow diagram'', showing trajectories of the system  in the plane $(|\alpha(t)|, |\beta(t)|)$ of local dynamical exponents for different values of $p_0$ around $p_c$. For $p_0>p_c$ (red curves), the system converges toward the thermal fixed point $(\alpha, \beta)=(0,0)$ (red star), whereas for $p_0<p_c$ (blue curves), it approaches the coarsening fixed point $(\alpha, \beta)=(3/2, 1/2)$ (blue star). For $p_0\leq1/\xi$ (darkest blue curve), the fixed point is reached directly, whereas for $p_0>1/\xi$ (lighter blue curves), the system first passes through a transient coarsening regime around $t=t^*$ (black dots). For $p=p_c$ (green curve), the system approaches the critical fixed point $(\alpha, \beta)=(1.5, 0.75)$ at long times (green star). In this case, the trajectory follows a straight line, reflecting the self-similar relation $\alpha(t)/\beta(t) = 2$. The blue curves correspond to $p_0\xi=0.2,1.26,1.58,2,2.51,3.16,7.64$, the green curve to $p_0\xi=7.71$, and the red curves to $p_0\xi=7.79,8.49,10$.
}
\end{figure*}

The collapse shown in the right panel of Fig. \ref{Fig:distributions}c further reveals that, at long times, the scaling function again decays algebraically at intermediate momenta, $f(x)\sim 1/x^\gamma$, but now with $\gamma=4$. 
In contrast to the weak-turbulence regime, here $\alpha/\beta=d\ne\gamma$, indicating that the dynamics is no longer self-similar. This signals a different microscopic mechanism: rather than being governed by weakly nonlinear wave interactions, the scaling dynamics at long times is controlled by the \emph{coarsening} of topological defects (vortices)  that emerge in the Bose gas beyond $t^*$.
To support this conjecture, we first note that the relation  $\alpha=d \beta$ implies that the correlation function $\langle\psi^\dagger(0,t)\psi(\br,t)\rangle=\text{FT}[n_\bp(t)](\br)=\tilde f(r/L(t))$ depends on a single characteristic length scale, $L(t)\propto t^\beta$, 
which we identify with the mean separation between defects. The microscopic picture is then the following. Beyond $t^*$, a  tangle of vortex lines forms in the gas. At later times, these vortices interact and reconnect, so that their mean separation $L(t)$ grows, as sketched in the insets of Fig.~\ref{Fig:distributions}c \cite{Nowak2012}.
The exponent  $\beta=1/2$ then follows from a simple dimensional argument: the velocity $v$ of a vortex line is set by its line tension $F$, which scales inversely with its radius of curvature. Dimensional analysis therefore gives $v\sim dL(t)/dt\sim F\sim 1/L(t)$, from which it follows that $L(t)\sim t^{1/2}$. 
This picture also provides a geometric interpretation of the $1/p^4$ decay of the scaling function. Consider two points separated by a distance $\xi\ll r \ll L$. Their correlation is approximately given by the probability that no vortex intersects the segment connecting them, which for small $r/L$ can be Taylor expanded as $\sim 1-r/L$  \cite{Bray2002}. 
Fourier transforming this form gives $\smash{n_\bp(t)\sim \int_\xi^L d^3\br e^{i\bp.\br}  (1-r/L)\sim 1/p^{d+1}}$ for $1/L\ll p\ll 1/\xi$, in agreement with the numerical results. 
Finally, we note that the momentum distribution at the coarsening NTFP exhibits a bimodal structure: intermediate momenta are governed by coarsening, $n_\bp(t)\sim p^{-4}$, whereas larger momenta retain the thermal scaling $n_\bp(t)\sim p^{-2}$, see Fig.~\ref{Fig:distributions}b. This bimodal structure is reminiscent of the condensate-plus-thermal distribution expected after equilibration of the Bose gas.

When $p_0<1/\xi$, corresponding to deep quenches that drive the Bose gas close to its ground state, Fig.~\ref{Fig:n0} shows that the weak-turbulence regime disappears entirely. The system then evolves directly toward the coarsening fixed point, without an intermediate stage, as further discussed in Appendix~\ref{app:strong_quench}. The corresponding dynamical phase diagram in Fig.~\ref{Fig:phasediag} summarizes the evolution as a function of time and quench depth. For $p_0<p_c$, the orange dotted curve marks the crossover time $t^*$ separating the weak-turbulence and coarsening fixed points, while the blue dashed curve indicates the time at which the dynamical exponent reaches $\alpha=3/2$ within $90\%$ accuracy. The phase diagram also reveals an optimal quench for accessing the coarsening fixed point, located around $p_0\sim1/\xi$.

\section*{Critical fixed point}

Finally, we consider the special case $p_0=p_c$ of a cooling quench precisely to the condensation threshold, Fig. \ref{Fig:distributionsCFP}a. At first sight, the zero-mode occupation $n_0(t)$ (green curve in Fig.~\ref{Fig:n0}) appears qualitatively indistinguishable from the case $p_0<p_c$: at long times, one again finds $n_0(t)\sim t^{3/2}$. A closer inspection of the full momentum distributions $n_\bp(t)$, see Figs. \ref{Fig:distributionsCFP}b-c, however, reveals a fundamentally different dynamical behavior. Indeed, for $p_0=p_c$, the distributions collapse almost perfectly when rescaled using the  exponents $\alpha=3/2$ and $\beta=3/4$.  The value $\beta=3/4$ is independently confirmed by an analysis of the cascade front presented in Appendix~\ref{app:front}.  The different value of $\beta$ demonstrates that the system approaches a distinct type of NTFP, which we refer to as the \emph{critical fixed point}. Moreover, at the critical fixed point the scaling function again exhibits an algebraic decay at intermediate momenta, $f(x)\sim x^{-\gamma}$, but now with $\gamma=2$. Remarkably, while the convergence of $\alpha$ and $\beta$ requires long evolution times, this algebraic decay already emerges on microscopic timescales $t\sim1/g\rho$, and persists throughout the evolution. 

To understand the origin of this critical dynamics, we note that, for $p_0=p_c$, the thermalization process should ultimately drive the system toward the quasi-long-range coherence
$\smash{\langle \psi^\dagger(0)\psi(\br)\rangle\sim 1/r}$ which characterizes equilibrium of 3D Bose gases at criticality \cite{Donner2007}. Out of equilibrium, this coherence develops progressively, spreading from microscopic  scales $\sim \xi$ to increasingly larger distances. At finite times, the corresponding momentum distribution is therefore expected to satisfy $\smash{n_\bp(t)= \int_{\xi}^{1/p_f(t)} d^3\br e^{i\bp.\br}  \langle \psi(0)\psi^*(\br)\rangle}\sim 1/p^2$ in the momentum range $p_f(t) \ll p\ll 1/\xi$, where $p_f(t)$ denotes the  correlation front. In contrast to quenches across the transition, whose late-time dynamics is governed by the diffusive coarsening of vortex lines, the critical quench is controlled by the spreading of critical fluctuations. At long times, this spreading is \emph{superdiffusive}, with $p_f(t)\propto t^{3/4}$, in sharp contrast to the diffusive growth of the coarsening fixed point. The critical fixed point also satisfies the exponent relation $\gamma=2=\alpha/\beta$. This relation implies a self-similar evolution of $n_\bp(t)$, which we interpret as the dynamical emergence of scale invariance, a defining hallmark of equilibrium critical phenomena. Remarkably, although the local exponents $\alpha(t),\beta(t)$  converge to their asymptotic values $(3/2,1/2)$ only at long times, the relation $\gamma=2=\alpha(t)/\beta(t)$ already holds on microscopic timescales, further supporting the emergence of scale invariance far from equilibrium.

The convergence of the post-quench dynamics toward the thermal, coarsening, and critical fixed points is summarized by the flow diagram in Fig.~\ref{Fig:distributionsCFP}d, which shows the trajectory of the system in the plane $(\alpha(t),\beta(t))$ of the  local exponents $\alpha(t)$ and $\beta(t)$, extracted respectively from the zero-momentum population (Fig.~\ref{Fig:n0}) and the cascade front (Fig.~\ref{Fig:front}). The trajectories clearly highlight the distinct convergence behaviors obtained as $p_0$ is varied around $p_0$. For the critical quench $p_0=p_c$, the trajectory reduces to a straight line, reflecting the scale-invariance relation $\alpha(t)/\beta(t)=2$.


\section*{Discussion and conclusion}
\label{Sec:discussion}

We have shown that cooling quenches across the Bose-Einstein condensation transition give rise to a  dynamical phase transition. Depending on the post-quench energy, the system is attracted toward distinct universal fixed points: a thermal fixed point above the transition, a succession of weak turbulence and a coarsening fixed point below it, and a critical fixed point for quenches to the threshold. In this picture, the equilibrium critical point does not merely mark the onset of condensation: it also separates distinct basins of attraction of the far-from-equilibrium dynamics. Table \ref{Table1} summarizes the main properties of the two NTFPs and transient weak turbulence studied in this work. Together, these results provide a unified picture in which equilibrium criticality shapes the universal long-time quench dynamics. 

At a theoretical level, this long-time dynamics cannot be captured by conventional Boltzmann kinetic theory. A non-perturbative description is essential to describe the dynamical condensation transition. In particular, below the transition, it regularizes the long-standing finite-time singularity of the Boltzmann equation, revealing the crossover from (perturbative) weak-wave turbulence to (non-perturbative) vortex-driven coarsening. At the condensation threshold, it further shows that the dynamics never enters the weak-turbulence regime but instead remains scale invariant throughout its evolution, exhibiting superdiffusive spreading of critical fluctuations rather than defect-driven coarsening.

The dynamical phase diagram uncovered here provides a natural framework for future studies of critical relaxation in isolated quantum gases. 
An immediate extension would be the investigation of two-time observables and aging phenomena at the critical fixed point, where universal aging exponents are expected to emerge, similarly to dynamical critical points previously identified, e.g.,  in classical critical quenches \cite{Janssen1989, Huse1989, Calabrese2005} or in quenched $\mathcal{O}(N)$ models \cite{Maraga2015}. Other important questions concern the role of quantum fluctuations in the dynamical phase diagram considered here, or the impact of trapping potentials in a cold-atom context. 
More broadly, our results establish Bose-Einstein condensation as a paradigmatic example of a phase transition possessing a complete far-from-equilibrium counterpart governed by universal dynamical scaling.

\begin{table}
\renewcommand{\arraystretch}{1.25} 
\setlength{\tabcolsep}{6.5pt}  
\begin{tabular}{ |c|c|c|c| } 
  \hline
   Dynamical  &  Weak turbulence & Coarsening & Critical\\ 
    exponent&  $1/\xi\leq p_0<p_c$ & $p_0<p_c$  & $p_0=p_c$ \\ 
  \hline\hline
  $\alpha$ & -2.7 & 3/2 & 3/2 \\ 
  $\beta$ & -1.1 & 1/2& 3/4 \\ 
  $\gamma$ & $\alpha/\beta=2.45$ & $d+1=4$ & $\alpha/\beta=2$\\\hline\hline
 $\text{Self-similar}$ & $\text{yes}$ & $\text{no}$ & $\text{yes}$\\
 $\text{Time range}$ & $t\sim t^*$ & $t\gg t^*$ & $t\gg 1/g\rho$\\
  \hline
\end{tabular} 
\caption{
\label{Table1}
Main properties of the dynamical regimes emerging from quenches below and to the condensation threshold.
}
\end{table}

\section*{Acknowledgements}
NC acknowledges  financial support of Agence Nationale de la Recherche (ANR), France, under the Grant No. ANR-24-CE30- 6695 FUSIoN. ASB acknowledges financial support from QuantEdu-France. The authors thank Quentin Glorieux, Ethan Uzan, Raul Teixeira, Leonardo Mazza, Nicolas Dupuis and Hadrien Kurkjian for useful exchanges and discussions.
\appendix

\section{Numerical resolution of the kinetic equation}
\label{app:num}

After a few particle-scattering events, the solution of the kinetic equation, Eq.~(\ref{Boltzmann_eq}), becomes isotropic, i.e., $n_\bp(t)=n_p(t)$. Under this assumption, the angular integrations in Eq.~\eqref{Boltzmann_eq} can be carried out analytically, yielding
\begin{align}\label{Boltzmann_isotropic}
&\frac{\partial n_p}{\partial t} = \frac{4\pi}{\mathcal{N}^2} \int_{p_1^2+p_2^2-p^2>0} \!\!\!\!\!\!\!\!\!\!\!\!\!\!\!\!\!\!\!\!\!\! dp_1 dp_2 \ p_1 p_2 W_\text{eff}(p, p_1, p_2, p_3)  \nonumber\\
&\ \ \ \ \times \left[n_{p_1}n_{p_2}(1\!+\!n_{p}\!+\!n_{p_3})\!-\!n_{p}n_{p_3}(1\!+\!n_{p_1}\!+\!n_{p_2})\right],
\end{align}
where $p_3\equiv \sqrt{p_1^2+p_2^2-p^2}$ and $\mathcal{N} \equiv 4\pi^2 \rho \xi^3 \gg 1$. Here and throughout this Appendix, we use dimensionless variables, with momenta measured in units of $1/\xi$ and times in units of $1/g\rho$. Introducing the variable $u\equiv \cos \theta$, with $\theta$ the angle between $\bp$ and $\bp_2$, the dimensionless interaction kernel reads
\begin{align}
\label{interaction_kernel}
   &W_\text{eff} (p, p_1, p_2, p_3) = \frac{p_2}{2} \int_{-1}^1 \frac{d u}{y_{p,p_2}(u)} g_\text{eff}^2(p, p_2, u)\times  \nonumber\\
& \left[\Theta\left(y_{p,p_2}(u)\!-\!|p_3\!-\!p_1|\right)\!-\!\Theta\left(y_{p,p_2}(u)\!-\!(p_3\!+\!p_1)\right)\right],
\end{align}
where $y_{p,p_2}(u) \equiv |\bp-\bp_2| = \sqrt{p^2 + p_2^2 - 2pp_2 u}$ and $\Theta$ denotes the Heaviside step function. The effective coupling, $g_\text{eff} (p, p_2, u) \equiv g_\text{eff} (p^2-p_2^2, |\bp-\bp_2|)$, is given by Eq.~\eqref{eq:geff}, with
\begin{align}\label{geff_num}
   &g\Pi^R (p^2-p_2^2, |\bp-\bp_2|)  = \frac{1}{2\mathcal{N}}\int \frac{k dk\, n_{k}}{y_{p,p_2}(u)}  \nonumber\\
   &\times \left[
   \ln\left( \frac{y_{p,p_2}^2(u)\! +\! |p^2\!-\!p_2^2| \!+ \!2 k y_{p,p_2}(u) \!+\! i0^+}{y_{p,p_2}^2(u)\! + \!|p^2\!-\!p_2^2|\! - \!2k y_{p,p_2}(u) \!+ \!i0^+} \right) \right.   \nonumber\\
   &\left.- \ln\left( \frac{-y_{p,p_2}^2(u) \!+\! |p^2\!-\!p_2^2|\! + \!2 k y_{p,p_2}(u) \!+ \!i0^+}{-y_{p,p_2}^2(u) \!+ \!|p^2\!-\!p_2^2| \!- \!2 k y_{p,p_2}(u)\! +\! i0^+} \right) \right].
\end{align}
For the numerical implementation, it is convenient to rewrite the interaction kernel as
\begin{align}
    W_\text{eff} (p, p_1, p_2, p_3) =\frac{p_2}{2} \int_{y_{p,p_2}(u) \in I} du  \, \frac{g_\text{eff}^2(p, p_2, u)}{y_{p,p_2}(u)}
\end{align}
where $I\equiv[\max(|p_3-p_1|, |p-p_2|), \min(p_3+p_1, p+p_2)]$. 
To evaluate the various integrals, momenta are discretized on a logarithmic grid (see below for details), extending from $p_\text{min}$ to $p_\text{max}$. These bounds are chosen to ensure that the relative deviations of both the total energy and the total particle number remain below $1\%$.
Throughout this work, we typically use $p_\text{min}=10^{-5}p_0$ to $10^{-3}p_0$ and $p_\text{max}=10^2p_0$. The different momentum integrals are evaluated using optimized quadrature schemes:
\begin{itemize}
    \item Evaluation of $g\Pi^R$. The integral is split into the three intervals $k \in [p_\text{min}, k^-] \cup [k^-, k^+] \cup [k^+, p_\text{max}]$, where
    \begin{align}
    k^{\pm} = \frac{1}{2} \left(y_{p,p_2}(u) \pm \frac{|p^2-p_2^2|}{y_{p,p_2}(u)} \right),
    \end{align}
    corresponding to the logarithmic singularities in Eq.~\eqref{geff_num}. Each interval is integrated using fourth-order sinh--tanh Gaussian-type quadratures.
    \item Evaluation of $W_\text{eff}$. The angular integral over $u\in[-1,1]$ is computed using a 150-point Gauss--Legendre quadrature. A mask is then applied to retain only the values satisfying  $y_{p,p_2}(u) \in [\max(|p_3-p_1|, |p-p_2|), \min(p_3+p_1, p+p_2)]$ for all $u$. 
    \item Evaluation of the collision integral. The $(p_1,p_2)$ integration domain is divided into the four regions $[p_\text{min}, p]\times[p_\text{min}, p] \cup [p, p_\text{max}]\times[p, p_\text{max}] \cup [p_\text{min}, p]\times[p, p_\text{max}] \cup [p, p_\text{max}]\times[p_\text{min}, p]$, which are integrated using exponential Gaussian-type quadratures with 300 points (numerical density parameter set to $-2$) on $[p_\text{min},p]$ and 100 points (numerical density parameter set to $5$) on $[p,p_\text{max}]$. These choices are motivated by the structure of the integrand in the $(p_1,p_2)$ plane.
\end{itemize}

The kinetic equation is solved for all $p\in[p_\text{min},p_\text{max}]$ on a grid designed to maximize the accuracy of the interpolation of $n_p$ while keeping the number of grid points as small as possible. We use a monotonic piecewise cubic Steffen interpolation for $n_p$, and the grid consists of the union of a logarithmically spaced grid of 500 points over $[p_\text{min},p_\text{max}]$ and an additional set of $3\times200=600$ points distributed over the intervals $[p_\text{min}, 0.01 p_0] \cup [0.01 p_0, 10 p_0] \cup [10 p_0, p_\text{max}]$, using exponential Gaussian-type quadratures to increase the point density in both the low-momentum range (where the inverse cascade develops) and the large-momentum range (tail of the distribution).

The numerical solution of the kinetic equation is implemented in Julia 1.12, while data analysis is performed in Python 3.11. Time integration is carried out using the Tsit5 solver from Julia's DifferentialEquations package, with an initial time step $\delta t=10^{-6}$, a relative tolerance of $10^{-3}$, and an absolute tolerance of $10^{-7}$. For the perturbative equation ($g_\text{eff}=g$), the blow-up time $t^*$ is estimated by fitting the long-time behavior of $n_0(t)$ to the scaling law \eqref{eq:WT_scaling} using the optimize.curve\_fit routine from Python's SciPy library.

\begin{figure}
\includegraphics[scale=0.37]{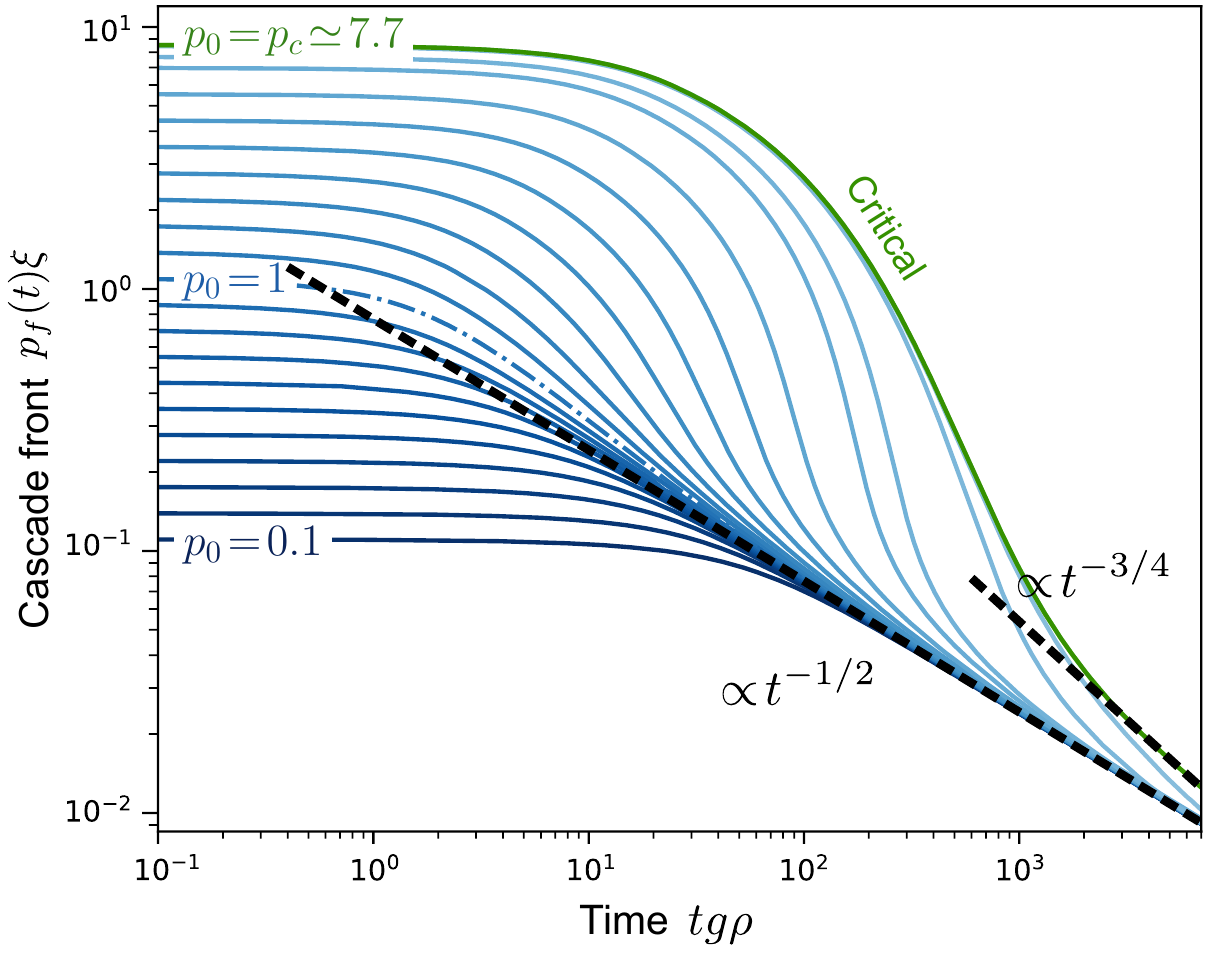}
\caption{
\label{Fig:front}
Inverse-cascade front momentum $p_f(t)$ as a function of time for increasing values of the post-quench momentum $p_0$, obtained by numerically solving  Eq.~\eqref{Boltzmann_eq} and using definition in Eq.~\eqref{eq:deffront}. 
For subcritical quenches ($p_0 < p_c$), the front momentum follows $p_f(t)\propto t^{-1/2}$ at long times. For weak quenches ($1/\xi\leq p_0<p_c$), the system first passes through a transient weak-turbulence regime in which $\smash{p_f(t)\propto |t-t^*|^{-\beta}}$ with $\beta\simeq -1.1$, before eventually crossing over to the long-time coarsening fixed point where $\beta=1/2$. For very deep quenches $p_0\leq1/\xi$ driving the Bose gas close to its ground state, the weak-turbulence regime disappears entirely. 
The dashed-dotted curve shows the front momentum evolution for $p_0=1/\xi$, for which the coarsening fixed point is reached most rapidly. At the critical quench ($p_0 = p_c$, green curve), the asymptotic scaling changes to $p_f(t)\propto t^{-3/4}$. 
In the figure, $p_0$ is expressed in units of $1/\xi=\sqrt{2mg\rho}$.
}
\end{figure}
\section{Inverse-cascade front}
\label{app:front}

While the dynamical exponent $\alpha$ is determined numerically from the time evolution of the zero-momentum occupation $n_0(t)$, the exponent $\beta$ characterizes the evolution of the inverse-cascade front momentum $p_f(t)$. In particular, the long-time scaling form in Eq.~\eqref{eq:WT_scaling} predicts
\begin{equation}
p_f(t) \sim t^{-\beta}.
\end{equation}
To extract $\beta$ numerically, we adopt the following systematic definition of $p_f(t)$. We fix a positive constant $\lambda\lesssim1$ and define $p_f(t)$ as the lowest momentum satisfying
\begin{align}
\label{eq:deffront}
    \left| \ln n_0(t) - \ln n_{p_f}(t) \right| \geq \lambda.
\end{align}
This criterion allows us to determine the front momentum at each time step. In our work, we take $\lambda=0.6$. The resulting evolution of $p_f(t)$ is shown in Fig.~\ref{Fig:front} for different values of the quench parameter $p_0$. 
The dynamics of $p_f(t)$ displays qualitatively distinct behaviors depending on whether the quench is subcritical ($p_0<p_c$) or critical ($p_0=p_c$). In both cases, the front momentum decays algebraically at long times according to $p_f(t)\propto t^{-\beta}$, but with $\beta\simeq 1/2$ for subcritical quenches and $\beta\simeq 3/4$ at the condensation threshold. 
For weak quenches below the threshold, $1/\xi\leq p_0<p_c$, the system first passes through a transient weak-turbulence regime in which $\smash{p_f(t)\propto |t-t^*|^{-\beta}}$ with $\beta\simeq -1.1$, before eventually crossing over to the asymptotic coarsening fixed point characterized by $\beta=1/2$.

\section{Effective coupling at long times}
\label{app:geff}

\begin{figure}
\includegraphics[scale=0.6]{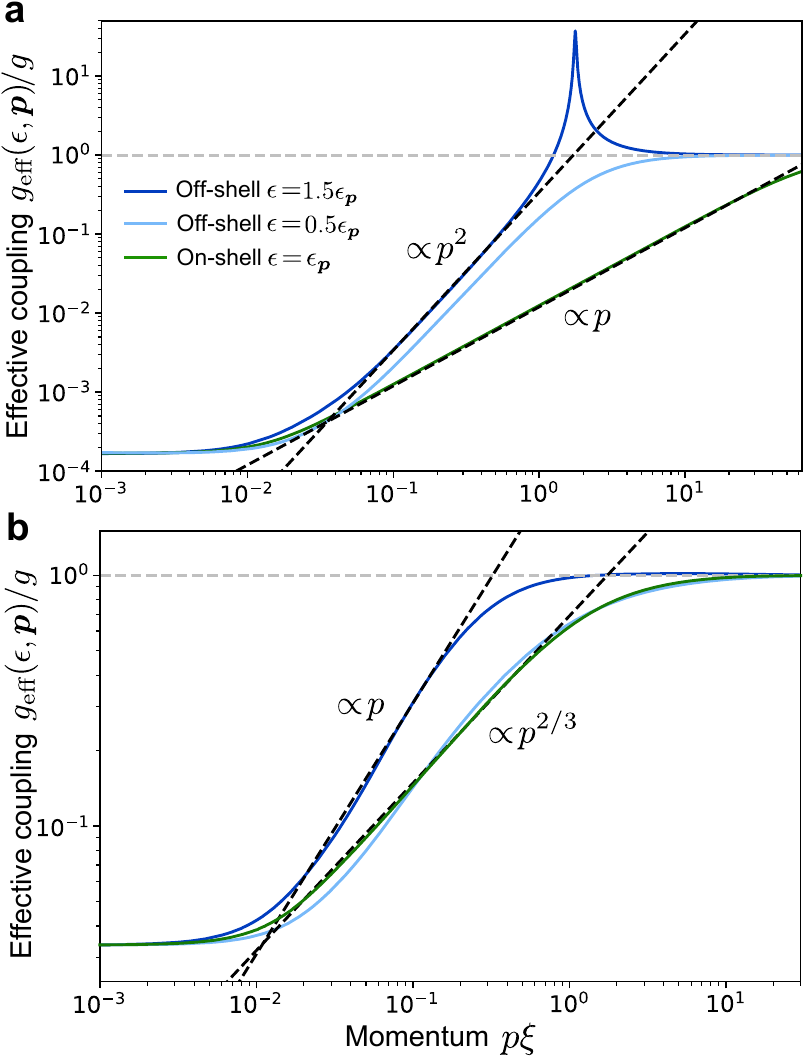}
\caption{
\label{Fig:Effective_coupling}
Effective coupling strength $g_\text{eff}(\epsilon, \bp)/g$, evaluated with the exact momentum distribution $n_\bp(t)$ computed numerically at long times by solving the kinetic equation \eqref{Boltzmann_eq}. Here time is fixed to $t = 7000/g\rho$. Panel \textbf{a} shows $g_\text{eff}(\epsilon, \bp)$ for $p_0/\xi=0.63$, corresponding to a deep quench below the critical point, and panel \textbf{b} shows $g_\text{eff}(\epsilon, \bp)$ for $p_0=p_c$, corresponding to a quench to the critical point. In each panel, the dark blue curve corresponds to $\epsilon=1.5\epsilon_\bp$, the light blue one to $\epsilon=0.5\epsilon_\bp$ (off-shell cases) and the green one to $\epsilon=\epsilon_\bp$ (on-shell case).}
\end{figure}

In addition to the momentum distribution, it is instructive to examine the structure of the effective coupling $g_{\mathrm{eff}}(\epsilon,\bp)$ during the evolution, particularly at long times, where it is expected to differ substantially from the Boltzmann approximation, $g_{\mathrm{eff}}=g$.
Fig.~\ref{Fig:Effective_coupling} shows  the ratio $g_{\mathrm{eff}}(\epsilon,\bp)/g$ at the fixed long time $t=7000/g\rho$ for two values of the post-quench momentum. Panel a corresponds to a deep quench below the critical point, $p_0=0.63/\xi$, while panel b corresponds to a quench precisely to the critical point, $p_0=p_c$. In each panel, we plot $g_{\mathrm{eff}}(\epsilon,\bp)$ for three representative energies: the on-shell value, $\epsilon=\epsilon_{\mathbf{p}}$, and two off-shell values, $\epsilon=1.5\epsilon_\bp$ and $\epsilon=0.5\epsilon_\bp$. We emphasize that both on-shell and off-shell processes contribute, in principle, to the collision integral in Eq.~\eqref{Boltzmann_eq}.

As is immediately apparent from Fig.~\ref{Fig:Effective_coupling}, the effective coupling exhibits several characteristic momentum regimes in which it follows distinct algebraic scalings,
\begin{align}
    g_\text{eff}(\epsilon, \bp) \propto \bp^z, \quad \quad z \in \{0 ; 2/3 ; 1 ; 2\},
\end{align}
where the exponent $z=2/3$ arises only for quenches to the critical point. For quenches across the transition, the three exponents $z=0$, $1$, and $2$, visible in Fig.~\ref{Fig:Effective_coupling}a, were previously predicted in Ref.~\cite{Chantesana2019} using a semi-analytical approach in which Eq.~\eqref{eq:geff} was evaluated with a trial momentum distribution exhibiting algebraic decay.

We now examine which of these scaling regimes is relevant for describing the long-time nonthermal fixed points observed numerically for the two types of quenches. We first consider quenches across the transition. In this case, the long-time inverse cascade is governed by particle-number conservation, which imposes the relation $\alpha/\beta=d=3$. Combined with the general exponent relation \eqref{eq:alphabeta_relation}, this constraint uniquely selects $z=2$ as the only exponent compatible with an inverse cascade, i.e., with $\alpha,\beta>0$. This implies that, for quenches across the transition, the long-time dynamics is dominated by off-shell contributions to the effective coupling.

For a critical quench, by contrast, the self-similar scaling of the momentum distribution imposes the relation $\alpha/\beta=2$ (see main text). Combined with the  exponent relation \eqref{eq:alphabeta_relation}, this uniquely selects $z=2/3$ as the only exponent compatible with an inverse cascade. In marked contrast to quenches across the transition, the long-time dynamics at criticality is therefore dominated by on-shell contributions to the effective coupling.

\begin{figure}
\includegraphics[scale=0.64]{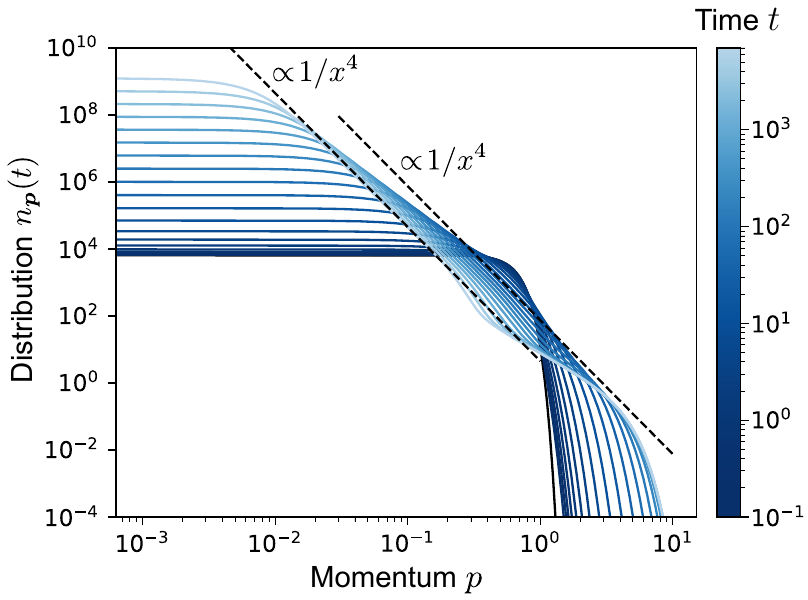}
\caption{
\label{Fig:Deep_quench}
Momentum distribution $n_\bp(t)$ at different times for a post-quench momentum $p_0=0.63/\xi$, corresponding to a deep quench across the transition. At intermediate momenta, the distribution rapidly develops the characteristic algebraic scaling $n_{\mathbf{p}}(t)\sim p^{-4}$ associated with vortex coarsening, with no signature of an intermediate weak-turbulence regime, in agreement with Fig.~\ref{Fig:n0}.
At large momenta, the distribution also exhibits a pronounced direct cascade. Throughout the figure, momenta are expressed in units of $1/\xi$ and times in units of $1/g\rho$.
}
\end{figure}
\section{Momentum distribution for strong quenches}
\label{app:strong_quench}

For deep quenches that drive the Bose gas close to its ground state, i.e., when the quench parameter satisfies $p_0<1/\xi$,  the system again evolves at long times toward a coarsening fixed point. Fig.~\ref{Fig:Deep_quench} shows the corresponding evolution of the momentum distribution. At long times, rescaling the distributions with the exponents $\alpha=1.5$ and $\beta=0.5$ yields the same data collapse as in Fig.~\ref{Fig:distributions}, demonstrating that the coarsening fixed point is qualitatively unchanged compared with the weak-quench regime.

The dynamics nevertheless differs from that of weak quenches in two important aspects. First, consistent with Fig.~\ref{Fig:n0}, no trace of a transient weak-turbulence regime is observed in the momentum distribution. Instead, the intermediate-momentum region exhibits the scaling $n_{\mathbf{p}}(t)\sim p^{-4}$ almost throughout the entire evolution. Second, a pronounced \emph{direct} cascade develops at large momenta. This cascade was predicted theoretically and observed in \textit{ab initio} simulations in Ref.~\cite{Chantesana2019}, and was subsequently observed experimentally in Ref.~\cite{Glidden2021}. It is also captured by the present quantum-kinetic framework and involves specific dynamical exponents. A dedicated study of direct cascades across the BEC transition will be investigated in detail in future work.


\begin{thebibliography}{99}

\bibitem{Polkovnikov2011}
A. Polkovnikov, K. Sengupta, A. Silva, and M. Vengalattore, 
\emph{Colloquium: Nonequilibrium dynamics of closed interacting quantum systems}, 
Rev. Mod. Phys. \textbf{83}, 863 (2011). 

\bibitem{Hohenberg1977}
 P. C. Hohenberg and B. I. Halperin, 
 \emph{Theory of dynamic critical phenomena}, 
 Rev. Mod. Phys. \textbf{49}, 435 (1977).
 
  \bibitem{Nazarenko2011}
 S. Nazarenko,
 \emph{Wave turbulence}, 
 Vol. 825, Springer Science \& Business Media, 2011.
 
 \bibitem{Cugliandolo2015}
L. F. Cugliandolo, 
\emph{Coarsening phenomena}, 
C. R. Phys. \textbf{16}, 257 (2015).

\bibitem{Bray2002}
A. J. Bray, 
\emph{Theory of phase-ordering kinetics}, 
Adv. Phys. \textbf{51}, 481 (2002).

 \bibitem{Berges2008}
J. Berges, A. Rothkopf, and J. Schmidt, 
\emph{Nonthermal fixed points: effective weak coupling for strongly correlated systems far from equilibrium}, 
Phys. Rev. Lett. 101, 041603 (2008).

\bibitem{Chantesana2019}
I. Chantesana, A. P. Orioli, M. Wouters, and T. Gasenzer,
\emph{Kinetic theory of nonthermal fixed points in a Bose gas},
Phys. Rev. A \textbf{99}, 043620 (2019).

\bibitem{Mikheev2019}
A. N. Mikheev, C.-M. Schmied, and T. Gasenzer,
\emph{Low-energy effective theory of non-thermal fixed points in a multicomponent Bose gas},
Phys. Rev. A \textbf{99}, 063622 (2019).

\bibitem{Gliott2024}
E. Gliott, A. Ran\c con, and N. Cherroret,
\emph{From inverse-cascade to sub-diffusive dynamic scaling in driven disordered Bose fluids},
Phys. Rev. Lett. \textbf{133}, 233403 (2024).
 
\bibitem{Zhu2023}
Y. Zhu, B. Semisalov, G. Krstulovic, S. Nazarenko,
\emph{Self-similar evolution of wave turbulence in Gross-Pitaevskii system},
Phys. Rev. E \textbf{108}, 064207 (2023).

\bibitem{Nowak2012}
B. Nowak, J. Schole, D. Sexty, and T. Gasenzer,
\emph{Nonthermal fixed points, vortex statistics, and superfluid turbulence in an ultracold Bose gas},
Phys. Rev. A \textbf{85}, 043627 (2012).

\bibitem{Comaron2019}
P. Comaron, F. Larcher, F. Dalfovo, and N. P. Proukakis,
\emph{Quench dynamics of an ultracold two-dimensional Bose gas},
Phys. Rev. A \textbf{100}, 033618 (2019).

 \bibitem{Marino2022}
J. Marino, M. Eckstein, M. S. Foster, and A. M. Rey, 
\emph{Dynamical phase transitions in the collisionless pre-thermal states of isolated quantum systems: theory and experiments},
Rep. Prog. Phys. \textbf{85} 116001 (2022).

\bibitem{Chandran2013}
A. Chandran, A. Nanduri, S. S. Gubser, and S. L. Sondhi, 
\emph{Equilibration and coarsening in the quantum ${O(N)}$ model at infinite ${N}$},
Phys. Rev. B \textbf{88}, 024306 (2013).

\bibitem{Sciolla2013}
B. Sciolla and G. Biroli,
\emph{Quantum quenches, dynamical transitions, and off-equilibrium quantum criticality},
Phys. Rev. B \textbf{88}, 201110 (2013).

\bibitem{Diehl2017}
A. Chiocchetta, A. Gambassi, S. Diehl, and J. Marino,
\emph{Dynamical Crossovers in Prethermal Critical States},
Phys. Rev. Lett. \textbf{118}, 135701 (2017).

\bibitem{Smacchia2015}
P. Smacchia, M. Knap, E. Demler, and A. Silva,
\emph{Exploring dynamical phase transitions and prethermalization with quantum noise of excitations},
Phys. Rev. B \textbf{91}, 205136 (2015).

\bibitem{Cherroret2024}
N. Cherroret
\emph{Dynamical phase transition of light in time-varying nonlinear dispersive media},
Phys. Rev. A \textbf{109}, 013519 (2024).

\bibitem{delPozo2026}
F. del Pozo, T. Morvan, I. Fr\'erot, and N. Cherroret,
\emph{Entanglement entropy across the dynamical phase transition in the quantum $\mathcal{O}(N)$ model},
arXiv:2605.23600 (2026).

\bibitem{Jian2019}
S.-K. Jian, S. Yin, and B. Swingle,
\emph{Universal Prethermal Dynamics in Gross-Neveu-Yukawa Criticality},
Phys. Rev. Lett. \textbf{123}, 170606 (2019).

\bibitem{Jian2021}
S. Yin  and S.-K. Jian,
\emph{Fermion-induced dynamical critical point},
Phys. Rev. B \textbf{103}, 125116 (2021).


 \bibitem{Langen2015}
 T. Langen, R. Geiger, and J. Schmiedmayer, 
 \emph{Ultracold atoms out of equilibrium},
 Annu. Rev. Condens. Matter Phys. \textbf{6}, 201 (2015).

\bibitem{Navon2019}
N. Navon, C. Eigen, J. Zhang, R. Lopes, A. L. Gaunt, K. Fujimoto, M. Tsubota, R. P. Smith, and Z. Hadzibabic,
\emph{Synthetic dissipation and cascade fluxes in a turbulent quantum gas},
Science \textbf{366}, 382 (2019).

\bibitem{Bagnato2022}
A. D. Garc\'ia-Orozco, L. Madeira, M. A. Moreno-Armijos, A. R. Fritsch, P. E. S. Tavares, P. C. M. Castilho, A. Cidrim, G. Roati, and V. S. Bagnato, 
\emph{Universal dynamics of a turbulent superfluid Bose gas}, 
Phys. Rev. A \textbf{106}, 023314 (2022).

\bibitem{Galka2022}
M. Ga\l ka, P. Christodoulou, M. Gazo, A. Karailiev, N. Dogra, J. Schmitt, and Z. Hadzibabic,
\emph{Emergence of Isotropy and Dynamic Scaling in 2D Wave Turbulence in a Homogeneous Bose Gas},
Phys. Rev. Lett. \textbf{129}, 190402 (2022).

\bibitem{Beugnon2017}
J. Beugnon and N. Navon,
\emph{Exploring the Kibble-Zurek mechanism with homogeneous Bose gases}, 
 J. Phys. B: At. Mol. Opt. Phys. \textbf{50}, 022002 (2017).
 
\bibitem{Keim2017}
 S. Deutschl\"ander, P. Dillmann, G. Maret, and P. Keim,
 \emph{Kibble-Zurek mechanism in colloidal monolayers}, 
Proc. Natl. Acad. Sci. \textbf{112}, 6925 (2015).

\bibitem{Kollar2011}
M. Kollar, F. A. Wolf, and M. Eckstein,
\emph{Generalized Gibbs ensemble prediction of prethermalization plateaus and their relation to nonthermal steady states in integrable systems},
Phys. Rev. B \textbf{84}, 054304 (2011).

\bibitem{Gring2012}
M. Gring, M. Kuhnert, T. Langen, T. Kitagawa, B. Rauer, M. Schreitl, I. Mazets, D. A. Smith, E. Demler, and J. Schmiedmayer, 
\emph{Relaxation and Prethermalization in an Isolated Quantum System}, 
Science \textbf{337}, 1318 (2012).

\bibitem{Natu2013}
S. S. Natu and E. J. Mueller,
\emph{Dynamics of correlations in a dilute Bose gas following an interaction quench},
Phys. Rev. A \textbf{87}, 053607 (2013).

\bibitem{Pietraszewicz2019}
J. Pietraszewicz, M. Stobi\'nska, and P. Deuar,
\emph{Correlation evolution in dilute Bose-Einstein condensates after quantum quenches},
Phys. Rev. A \textbf{99}, 023620 (2019).

\bibitem{Larre2018}
P.-\'E. Larr\'e, D. Delande, and N. Cherroret,
\textit{Postquench prethermalization in a disordered quantum fluid of light},
Phys. Rev. A \textbf{97}, 043805 (2018).

\bibitem{Bardon-brun2020}
T. Bardon-brun, S. Pigeon, and N. Cherroret, 
\emph{Classical Casimir force from a quasi-condensate of light}, 
Phys. Rev. Research \textbf{2}, 013297 (2020).

\bibitem{Mallayya2019}
K. Mallayya, M. Rigol, and W. De Roeck,
\emph{Prethermalization and Thermalization in Isolated Quantum Systems},
Phys. Rev. X \textbf{9}, 021027 (2019).

\bibitem{Bertini2015}
B. Bertini, F. H. L. Essler, S. Groha, and N. J. Robinson,
\emph{ Prethermalization and Thermalization in Models with Weak Integrability Breaking},
 Phys. Rev. Lett. \textbf{115}, 180601 (2015).

\bibitem{Buchhold2016}
M. Buchhold, M. Heyl, and S. Diehl,
\emph{Prethermalization and thermalization of a quenched interacting Luttinger liquid},
Phys. Rev. A \textbf{94}, 013601 (2016).

 \bibitem{Regemortel2018}
M. V. Regemortel, H. Kurkjian, M. Wouters, and I. Carusotto,
\emph{Prethermalization to thermalization crossover in a dilute Bose gas following an interaction ramp},
Phys. Rev. A \textbf{98}, 053612 (2018).

\bibitem{Mori2018}
T. Mori, T. N Ikeda, E. Kaminishi, and M. Ueda,
\emph{Thermalization and prethermalization in isolated quantum systems: a theoretical overview},
 J. Phys. B: At. Mol. Opt. Phys. \textbf{51} 112001 (2018).

\bibitem{Duval2023}
C. Duval and N. Cherroret
\emph{Quantum kinetics of quenched two-dimensional {B}ose superfluids},
Phys. Rev. A \textbf{107}, 043305 (2023).

\bibitem{Duval2025}
C. Duval and N. Cherroret
\emph{Anomalous Landau damping and algebraic thermalization in two-dimensional superfluids far from equilibrium},
Phys. Rev. A \textbf{111}, L021301 (2025).

 \bibitem{Huh2024}
S. Huh,  K. Mukherjee, K. Kwon, J. Seo,  J. Hur, S. I. Mistakidis, H. R. Sadeghpour,  and J. Choi,
 \emph{Universality class of a spinor {B}ose-Einstein condensate far from equilibrium}
Nature Physics \textbf{20}, 402 (2024).

 \bibitem{Manovitz2025}
T.  Manovitz,  S. H. Li, S. Ebadi, R. Samajdar, A. A. Geim, S. J. Evered, D. Bluvstein, H. Zhou, N. Koyluoglu,  J. Feldmeier, P. E. Dolgirev, N. Maskara, M. Kalinowski, S. Sachdev, D. A. Huse, M. Greiner, V. Vuleti\'c, and M. D. Lukin.
 \emph{Quantum coarsening and collective dynamics on a programmable simulator},
Nature \textbf{638}, 86 (2025).

 \bibitem{Erne2018}
S. Erne,  R. B{\"u}cker, T. Gasenzer, J. Berges, and J. Schmiedmayer, 
\emph{Universal dynamics in an isolated one-dimensional {B}ose gas far from equilibrium},
Nature \textbf{563}, 225 (2018).

\bibitem{Glidden2021}
J. A. P. Glidden, C. Eigen, L. H. Dogra, T. A. Hilker, T. P. Smith, and Z. Hadzibabic ,
\emph{Bidirectional dynamic scaling in an isolated Bose gas far from equilibrium},
Nature Phys. \textbf{17}, 457 (2021).

\bibitem{Sunami2022}
S. Sunami, V. P. Singh, D. Garrick, A. Beregi, A. J. Barker, K. Luksch, E. Bentine, L. Mathey, and C. J. Foot,
\emph{Universal Scaling of the Dynamic BKT Transition in Quenched 2D Bose Gases},
Science \textbf{382}, 443 (2023).

\bibitem{Abuzarli2022}
M. Abuzarli, N. Cherroret, T. Bienaim\'e, and Q. Glorieux,
\emph{Non-equilibrium pre-thermal states in a two-dimensional photon fluid},
Phys. Rev. Lett. \textbf{129}, 100602 (2022).

\bibitem{Gazo2023}
M. Gazo, A. Karailiev, T. Satoor, C. Eigen, M. Ga\l ka, and Z. Hadzibabic,
\emph{Universal Coarsening in a Homogeneous Two-Dimensional Bose Gas},
Science \textbf{389}, 802 (2025).

\bibitem{Scoquart2022}
T. Scoquart, D. Delande, N. Cherroret,  
\emph{Dynamical emergence of a Kosterlitz-Thouless transition  in a disordered Bose gas following a quench}, 
Phys. Rev. A \textbf{106}, L021301 (2022).

\bibitem{Gliott2025}
E. Gliott, C. Piekarski, and N. Cherroret,
\emph{Coarsening of binary Bose superfluids: An effective theory},
Phys. Rev. Research \textbf{7}, 033189 (2025).

\bibitem{Griffin2009}
A. Griffin, T. Nikuni, and E. Zaremba, 
\emph{Bose-Condensed Gases at Finite Temperatures}, 
(Cambridge University Press, New York, 2009).
 
 \bibitem{Semikoz1995}
D. Semikoz, and I. Tkachev, 
\emph{Kinetics of Bose Condensation},
Phys. Rev. Lett. \textbf{74}, 3093 (1995).

\bibitem{Semikoz1997}
D. Semikoz, and I. Tkachev, 
\emph{Condensation of bosons in the kinetic regime},
Phys. Rev. D \textbf{55}, 489 (1997).

\bibitem{Berloff2002}
N. Berloff, B. Svistunov, 
\emph{Scenario of strongly nonequilibrated Bose-Einstein condensation},
Phys. Rev. A \textbf{66} 013603 (2002).

\bibitem{Josserand2006}
C. Josserand, Y. Pomeau, S. Rica, 
\emph{Self-similar Singularities in the Kinetics of Condensation},
Low Temp Phys \textbf{145}, 231 (2006).

\bibitem{Semisalov2021}
B. Semisalov, V. Grebenev, S. Medvedev, and S. Nazarenko, 
\emph{Numerical analysis of a self-similar turbulent flow in Bose-Einstein condensates},
Communications in Nonlinear Science and Numerical Simulation \textbf{102}, 105903 (2021).
 
\bibitem{Stringari_pitaevskii2003}
L. P. Pitaevskii and  S. Stringari,
\emph{Bose-Einstein condensation},
Oxford University Press (2003).

\bibitem{Smith2017}
R. P. Smith, 
\emph{Effects of Interactions on Bose-Einstein Condensation, in Universal Themes of Bose-Einstein}, 
eds N. Proukakis, D. Snoke, and P. Littlewood (Cambridge UniversityPress, Cambridge, 2017).

\bibitem{Donner2007}
T. Donner, S. Ritter, T. Bourdel, A. \"{O}ttl, M. K\"{o}hl, and  T. Esslinger,
\emph{Critical Behavior of a Trapped Interacting Bose Gas},
 Science \textbf{315}, 1556 (2007).

\bibitem{footnote1}
Note that to achieve this collapse, we include the time scale $t^*$ in the rescaling procedure, although its effect becomes negligible for $t\gg t^*$.  The intermediate regime where $t^*$ remains relevant was referred to as a pre-scaling stage in \cite{Schmied2019, Gazo2023}. 

 
\bibitem{Schmied2019}
C.-M. Schmied, A. N. Mikheev, and T. Gasenzer, 
\emph{Prescaling in a Far-from-Equilibrium Bose Gas}, 
Phys. Rev. Lett. \textbf{122}, 170404 (2019).

\bibitem{Janssen1989}
H. K. Janssen, B. Schaub, B. Schmittmann,
\emph{New universal short-time scaling behaviour of critical relaxation processes},
Zeitschrift f\"ur Physik B \textbf{73}, 539 (1989). 

\bibitem{Huse1989}
D. A. Huse, 
\emph{Remanent Magnetization Decay at the Spin-Glass Critical Point: A New Dynamic Critical Exponent for Nonequilibrium Autocorrelations},
Phys. Rev. B \textbf{40}, 304 (1989).

\bibitem{Berthier2001}
L. Berthier, P. C. W. Holdsworth, and M. Sellitto,
\emph{Nonequilibrium critical dynamics of the two-dimensional XY model},
Journal of Physics A: Mathematical and General, \textbf{34}, 1805 (2001).


\bibitem{Calabrese2005}
P. Calabrese and A. Gambassi,
\emph{Ageing properties of critical systems},
Journal of Physics A \textbf{38}, R133 (2005).

\bibitem{Maraga2015}
A. Maraga, A. Chiocchetta, A. Mitra, and A. Gambassi,
\emph{Aging and coarsening in isolated quantum systems after a quench: Exact results for the quantum $\mathcal{O}(N)$ model with $N\to\infty$},
Phys. Rev. E \textbf{92}, 042151 (2015).









 



\end{thebibliography}
\end{document}